\documentclass[lettersize,journal]{IEEEtran}
\usepackage{xcolor,soul,framed}
\usepackage[pdftex]{graphicx}
\graphicspath{{../pdf/}{../jpeg/}}
\DeclareGraphicsExtensions{.pdf,.jpeg,.png}
\usepackage[cmex10]{amsmath}
\usepackage{amssymb,amsfonts}
\usepackage{graphicx}
\usepackage{enumitem}
\usepackage{textcomp}
\usepackage{colortbl}
\usepackage{color}
\usepackage{enumerate}
\usepackage{siunitx}
\usepackage{tabularx}
\usepackage{booktabs}
\usepackage{caption}
\usepackage{epstopdf}
\usepackage{cite}
\usepackage{subcaption}
\usepackage{float,multirow}
\usepackage{amsfonts}
\usepackage[none]{hyphenat}
\usepackage{url}

\usepackage[breaklinks]{hyperref}
\usepackage{comment}
\usepackage[euler]{textgreek}
\usepackage{epsfig,bm,amssymb,setspace}
\usepackage{fontawesome}
\usepackage[table,xcdraw]{xcolor}
\usepackage{listings}
\usepackage{algorithm}
\usepackage{threeparttable}
\usepackage[ruled,vlined,algo2e]{algorithm2e}
\usepackage[T1]{fontenc}
\usepackage{amsthm}
\usepackage{array}
\usepackage{balance}
\usepackage{pifont}
\usepackage{multirow}
\usepackage{algorithmic}
\usepackage{csquotes}
\usepackage{dblfloatfix}

\definecolor{darkgreen}{rgb}{0.01, 0.75, 0.24}
\definecolor{codeblue}{rgb}{0,0,0.7}
\definecolor{codegreen}{rgb}{0,0.5,0}
\definecolor{codegray}{rgb}{0.4,0.4,0.4}

\lstdefinestyle{verilogstyle}{
  language=Verilog,
  basicstyle=\ttfamily\scriptsize,
  keywordstyle=\color{codeblue}\bfseries,
  commentstyle=\color{codegreen}\itshape,
  numbers=none,
  frame=single,
  breaklines=true,
  showstringspaces=false,
  tabsize=2
}

\begin{document}

\title{Benchmarking LLMs for Verilog Design Flows}

\author{Angshuman~Chakravertty,~Rahul~Koshti,~Buddhi~Prakash~Sharma,~\IEEEmembership{Student Member,~IEEE,}
and~Vinay~Chamola,~\IEEEmembership{Senior Member,~IEEE}%
\thanks{Angshuman Chakravertty and Rahul Koshti are with the School of Technology Management and Engineering,
NMIMS, Hyderabad, India (e-mail: angshumanchakravertty2@gmail.com, rahul.koshti@nmims.edu).}%
\thanks{Buddhi Prakash Sharma is with the Department of Electronics and Communication Engineering,
Indian Institute of Technology, Roorkee, India (e-mail: coolbp.buddhiprakash@gmail.com).}%
\thanks{Vinay Chamola is with the Department of Electrical and Electronics Engineering and the Anuradha
and Prashanth Palakurthi Centre for Artificial Intelligence Research (APPCAIR), Birla Institute of
Technology and Science, Pilani (BITS Pilani), Pilani 333031, India
(e-mail: vinay.chamola@pilani.bits-pilani.ac.in).}}

\markboth{IEEE Design \& Test, VOL.~XX, NO.~XX, XX~2026}%
{Chakravertty \MakeLowercase{\textit{et al.}}: Benchmarking LLMs for Verilog Design Flows}

\maketitle

\begin{abstract}
Large language models (LLMs) show promise in code generation but their capabilities to produce correct, synthesizable hardware description language (HDL) code still remain to be properly benchmarked. Existing evaluations are primarily relying on pass@k metrics and lack proper end-to-end toolchain validation. This paper presents a reproducible benchmarking platform that evaluates open-source LLMs on Verilog RTL generation across 50 curated tasks consisting of combinational, sequential, finite state machine (FSM), and mixed designs. The pipeline consisting of constrained prompting, post-processing, and semantic-aware iterative refinement with waveform analysis, formal equivalence verification, and Abstract Syntax Tree (AST)-based repair validates the generated code via Verilator compilation and Icarus Verilog simulation. Across the 12 benchmarks and the 1,610 total runs evaluating three models of different sizes (Llama-3-8B, StarCoder2-7B, and TinyLlama-1.1B), the pipeline improved syntax validity from 0\% to a 70.43\% average and simulation pass rate to 51.8\% across three open-source models. Most notably TinyLlama (1.1B parameters) achieved the highest individual syntax validity at 80.0\%, with functional correctness comparable to the 8B model. The platform and dataset are open-source, enabling reproducible evaluation of generative AI for hardware design workflows. 
\end{abstract}


\section{Introduction}

\IEEEPARstart{T}{he} EDA industry has had a structural problem where the SoC complexity keeps growing, but the pool
of expert RTL verification engineers does not. LLMs have shown potential to help with HDL code generation, but widespread adoption raises one question that most published evaluations still tend to ignore: can the code actually be built?
Previously, Pass@k, a tool borrowed from software benchmarking, asked whether under some broad criterion, at least one of $k$ samples compiles or not, but this does not tell us whether the Verilog survives a real lint check, passes functional
simulation against a self-checking testbench, or synthesizes to logic gates. We measured all three, reproducibly, on open-source models without fine-tuning.
Table~\ref{tab:related} shows how this platform is different from any prior work, and the main distinction of this work is the depth of the toolchain and the repair coverage; prior efforts validate the generation quality with
simulation alone and generally skip iterative repair \cite{b2}. There are three results from this study that are to be flagged immediately: first, the constrained prompting alongside structured post-processing together lifted the syntax validity from 0\% to 60\%, showcasing a combined gain that is larger than any model-level effect. Second, FSM and mixed designs were both hard throughout, and none of the open-source models (small, medium, or large) reliably handle complex state machines without explicit structural scaffolding. Third result, TinyLlama-1.1B despite its releatively less parameter count led on syntax validity and matched Llama-3-8B closely on simulation pass rate, suggesting that workflow design can matter as much as model scale for constrained RTL generation tasks.

\begin{table*}[!t]
\centering
\caption{Comparison With Related LLM-Verilog Work Across Nine Evaluation Dimensions}
\label{tab:related}
\scriptsize
\setlength{\tabcolsep}{3pt}
\resizebox{\textwidth}{!}{%
\begin{tabular}{|l|c|c|c|c|c|c|c|c|c|}
\hline
\textbf{Work} & \textbf{Tasks} & \textbf{Categories} & \textbf{Models} &
\textbf{Metrics} & \textbf{Sim} & \textbf{Formal} & \textbf{Repair} &
\textbf{Synth/PPA} & \textbf{OSS} \\
\hline
VeriGen \cite{b1}      & 156    & C, S         & 5+  & pass@k          & Y & N & N       & N & Y* \\
VerilogEval \cite{b6}  & 156    & C, S         & 5+  & pass@k          & Y & N & N       & N & Y  \\
AutoVCoder \cite{b4}   & 156    & C, S         & 3   & pass@1/5/10     & Y & N & N       & N & N  \\
AutoBench \cite{b5}    & Custom & Various      & 3   & TB quality       & Y & N & Partial & N & N  \\
HaVen \cite{b7}        & 156    & C, S         & 3   & pass@1          & Y & N & N       & N & N  \\
\textbf{This work}     & \textbf{50} & \textbf{C, S, FSM, M} & \textbf{3} &
\textbf{Lint + sim + synth} & \textbf{Y} & \textbf{Y} & \textbf{Y} & \textbf{Y} & \textbf{Y} \\
\hline
\multicolumn{10}{l}{\scriptsize C=Combinational, S=Sequential, FSM=Finite State Machine,
M=Mixed; *=dataset only; OSS=Open-Source} \\
\end{tabular}}
\end{table*}
The platform and full dataset are publicly available at
\url{https://github.com/Ganglet/Benchmarking-Generative-AI-in-EDA-Workflows}.

\begin{figure*}[!t]
\centering
\includegraphics[width=0.95\linewidth]{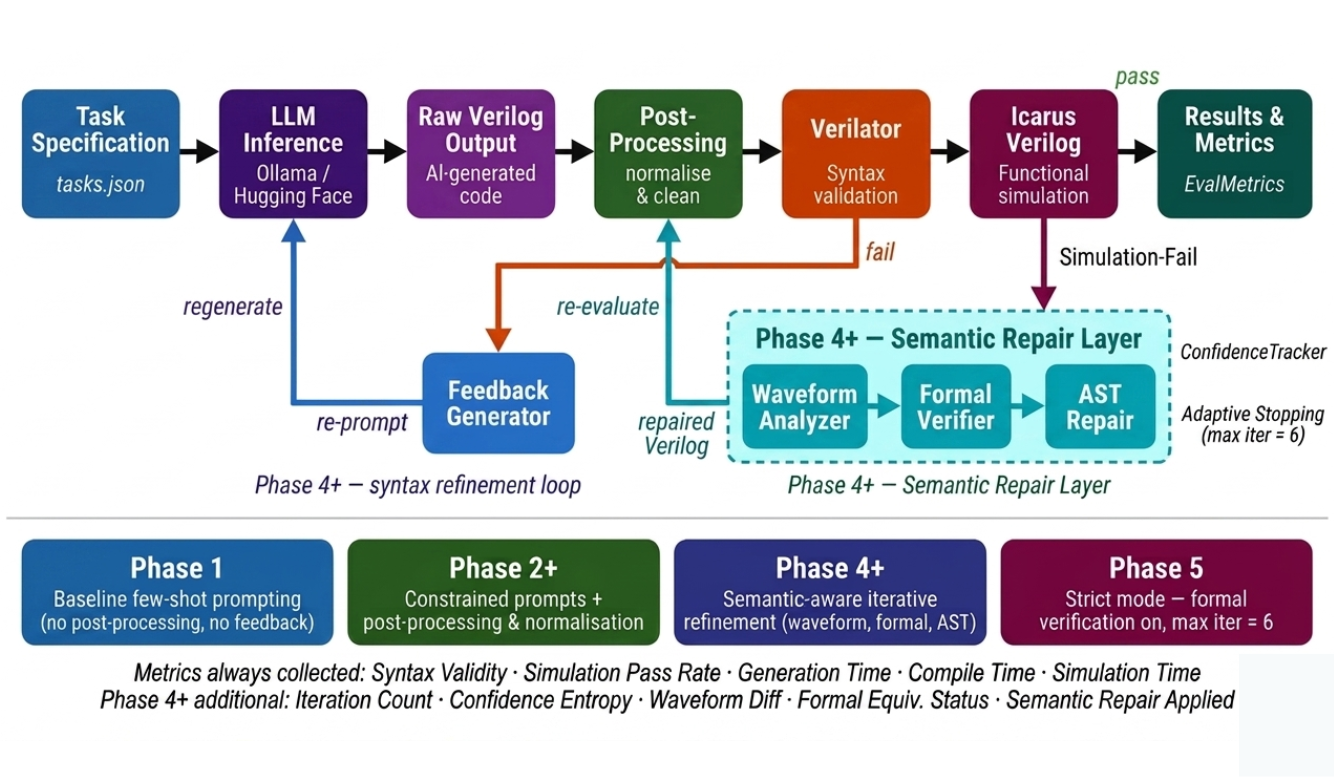}
\caption{Guarded generation pipeline architecture showing the five-phase evolution and the semantic repair layer.}
\label{fig:pipeline}
\end{figure*}
\section{Tutorial: A Safe LLM-to-Verilog Workflow}
In addition to defining a useful workflow pattern that any EDA team can follow even in the absence of the platform itself, there are four steps that must be taken in order to use this platform.

\textbf{Step 1: Specify before generating.} First, we need to
gather the expected module name and full port list (signal name, direction, and width), along with any structural constraints (clock domain, reset polarity, and FSM
topology) before touching the model. Our platform uses \texttt{tasks.json} to store it, but in practice, the
same can be done with a design spec or IP block requirement
document. Benchmark 1, indicated that most structural failures could be traced directly to the missing or ambiguous interface information and not model weakness.

\textbf{Step 2: Inject constraints into the prompt.} The precise module signature, the explicit Verilog-2001 rules (``only \texttt{wire} and \texttt{reg} are to be used and not \texttt{always\_ff}, \texttt{logic}, or \texttt{always\_comb}''), and a brief, accurate module example should all be included in the prompt. After which it is to be run at temperature 0.0, and before any toolchain check, post-process the
output with the objective of stripping BlueSpec artifacts, converting SystemVerilog keywords, and enforcing the declared port list.
Post-processing is an important process, as it converts a plausible-looking
output into something that the compiler will accept.

\textbf{Step 3: Validate with real tools.} At this stage, we have to run the Verilator in lint-only mode, as this catches any syntax errors, undeclared signals, and port mismatches, and if it passes, the DUT should be paired with a self-checking testbench and then the Icarus Verilog simulation should be run with a 30-second timeout at both stages. Yosys generic synthesis is the final gate for RTL bound for silicon as it confirms the output maps to the logic without any synthesis-blocking constructs.

\textbf{Step 4: Repair with targeted feedback.} In last stage, when failure occurs, distill the error into a compact repair prompt under 500 characters and regenerate; this should be repeated up to five times. For FSM and complex
tasks, the supplement simulation feedback with waveform-level mismatch analysis and Yosys formal equivalence checking exposes errors that the testbench stimulus vectors alone might miss.

These four stages sequence treats LLM-generated RTL as a draft that needs validation and repair, not a deliverable.

\section{Benchmark Design and Dataset}

There are 50 jobs in the dataset: 23 combinational, 14 sequential, 8 FSM, and 5 mixed designs. Gates, adders, multiplexers, and decoders are examples of combinational tasks that assess Boolean logic without taking temporal behavior into account. Clock-dependent operation is introduced by the sequential activities, such as flip-flops, counters, and shift registers. For the FSM tasks, like sequence detector, traffic light controller and turnstile controller etc., the model must consistently encode states during a series of transitions as well as generating outputs correctly. The mixed tasks (e.g., priority encoders and ALUs) check the interaction between control and datapath logic over multiple \texttt{always} blocks.

Fifty tasks were chosen since each benchmark would result in 450 runs if three models were run three times and every run were then fully validated against a reference testbench, thus scaling further without fixed, deterministic testbenches, would instead trade reproducibility for coverage.
Each task further includes a synthesizable Verilog-2001 reference implementation, a self-checking testbench that applies deterministic stimuli along with counting pass/fail cases, and explicit I/O metadata in
\texttt{tasks.json}. Specification variants such as task IDs ending in \texttt{\_001}, \texttt{\_002}, \texttt{\_003} keep the testbench and reference constant while changing the natural language phrasing to isolate phrasing effects.

\begin{table*}[!t]
\centering
\caption{Benchmark Progression Across Pipeline, Model, and Configuration}
\label{tab:benchmark_metrics}
\small
\begin{tabular}{|p{0.8cm}|p{1cm}|p{1.7cm}|p{1.8cm}|p{1.8cm}|p{1.8cm}|p{1.5cm}|p{4cm}|}
\hline
\textbf{BM} & \textbf{Models} & \textbf{Syntax (\%)} & \textbf{Sim Pass (\%)} &
\textbf{Gen Time (s)} & \textbf{Compile (s)} & \textbf{Sim (s)} & \textbf{Key addition} \\
\hline
1  & 2 & 0.0   & 0.0  & ---        & ---   & ---   & Baseline few-shot \\
2  & 2 & 20.0  & 0.0  & ---        & ---   & ---   & Constrained prompt \\
3  & 2 & 60.0  & 30.0 & $\sim$3.4  & ---   & ---   & Post-processing \\
4  & 2 & 63.0  & 56.7 & 4.32       & 0.12  & 0.065 & Timing instrumentation \\
5  & 3 & 51.1  & 42.2 & 3.54       & 0.112 & 0.061 & Third model added \\
6  & 3 & 80.0  & 73.3 & 20.84      & 0.516 & 0.109 & Sequential normalization \\
7  & 3 & 29.4  & 20.0 & 4.38       & 0.122 & 0.033 & Scale to 20 tasks (FSM/mixed) \\
8  & 3 & 63.3  & 46.1 & 4.44       & 0.109 & 0.064 & FSM template scaffolding \\
9  & 3 & 67.8  & 48.9 & 3.56       & 0.136 & 0.076 & Iterative refinement (Phase 4) \\
10 & 3 & 69.6  & 48.7 & 5.11       & 0.211 & 0.104 & Scale to 50 tasks \\
11 & 1 & 71.3  & 61.3 & 4.77       & 0.174 & 0.084 & Generation time optimization \\
12 & 3 & 70.43 & 51.8 & 3.46       & 0.125 & 0.073 & Phase 5 FSM enhancements \\
\hline
\end{tabular}
\end{table*}
\section{Guarded Generation Pipeline}

This pipeline has evolved through five incremental phases, where each phase introduces an additional component, and the resulting performance variations form the basis of the ablation study, which could be seen in Fig.~\ref{fig:pipeline}.

\subsection{Phase 1: Baseline}
In phase 1, the baseline was set with vanilla few-shot prompting, no constraints, no post-processing, and no repair loop, and thus the syntax validity
at benchmark 1 was 0\%. Models generated the wrong module names with missing or mismatched port declarations, SystemVerilog constructs where Verilog-2001 was required, and (from TinyLlama especially) BlueSpec artifacts; the logic was occasionally identifiable, but the interface was never.

\subsection{Phase 2 and phase 3: Constrained Prompting and Post-Processing}
For phases 2 and 3, three new items were added to the prompt: (1) the exact expected module name, which was extracted from the task ID, (2) the full port specification from \texttt{tasks.json}, and (3) the Verilog-2001 syntax constraints. The post-processing (\texttt{post\_process\_verilog()}) then applies six sequential transforms: module name correction, port enforcement, \texttt{logic}$\rightarrow$\texttt{wire/reg} conversion, BSV artifact removal, sequential normalization (added in Phase 3), and FSM/mixed template scaffolding
(Phase 4). Together these have helped in lifting the syntax validity from 0\% to 55--93\% depending on the category, making the structure fixable, but the logic still remained wrong.

\subsection{Phase 4: Semantic-Aware Iterative Refinement}
In phase 4 a new generate$\rightarrow$evaluate$\rightarrow$feedback$\rightarrow$regenerate loop gets introduced, which runs for up to five iterations and is inspired by compiler-feedback repair approaches~\cite{b9}. The \texttt{FeedbackGenerator} now feeds the compiler errors and testbench mismatches back into the model as a targeted prompt capped at 500 characters, and this loop is further enhanced by three semantic techniques. To produce VCD files, \texttt{WaveformAnalyzer}
simulates both reference and generated designs and then compares the signal values cycle-by-cycle and flags the first mismatch; \texttt{FormalVerifier} runs Yosys \texttt{equiv\_make}/\texttt{equiv\_simple} with the objective of catching logic errors that stimulus-limited testbenches might have missed~\cite{b10}; \texttt{ASTRepair} then parses the generated Verilog using \texttt{pyverilog} and
applies structural fixes (like missing port declarations, undeclared signals) that regex post-processing cannot reach; \texttt{ConfidenceTracker} computes Shannon entropy of token log-probabilities;
if entropy is high, the expensive waveform and formal steps are skipped. Formal verification was
disabled by default in Phase 4 and re-enabled in Phase 5.

Tasks are classified into three tiers to allocate compute proportionally: Tier~0 (trivial
combinational, single iteration, no semantic tools), Tier~1 (standard sequential, up to five
iterations with waveform analysis and AST repair), and Tier~2 (complex FSM and mixed, up to
five iterations with the full semantic stack including formal equivalence checking).

\subsection{Phase 5: FSM and Mixed Enhancements}
Even with Phase 4 repair, FSM functional correctness stayed near zero. Four targeted
additions addressed this: (1) an explicit state transition table embedded in the prompt
(replacing natural language state descriptions); (2) contrastive broken/fixed FSM examples
showing common mistakes, including missing \texttt{default} cases and collapsed state spaces, alongside
corrections; (3) strict mode that disables entropy gating so every FSM task receives the full
semantic repair stack regardless of model confidence; and (4) a micro-repair pass targeting
missing \texttt{default} statements, incomplete \texttt{begin/end} blocks, and wrong reset
logic. The iteration cap increased from 5 to 6; the improvement threshold dropped from 0.1
to 0.05.

\begin{figure*}[!ht]
    \centering
    \includegraphics[width=1\linewidth]{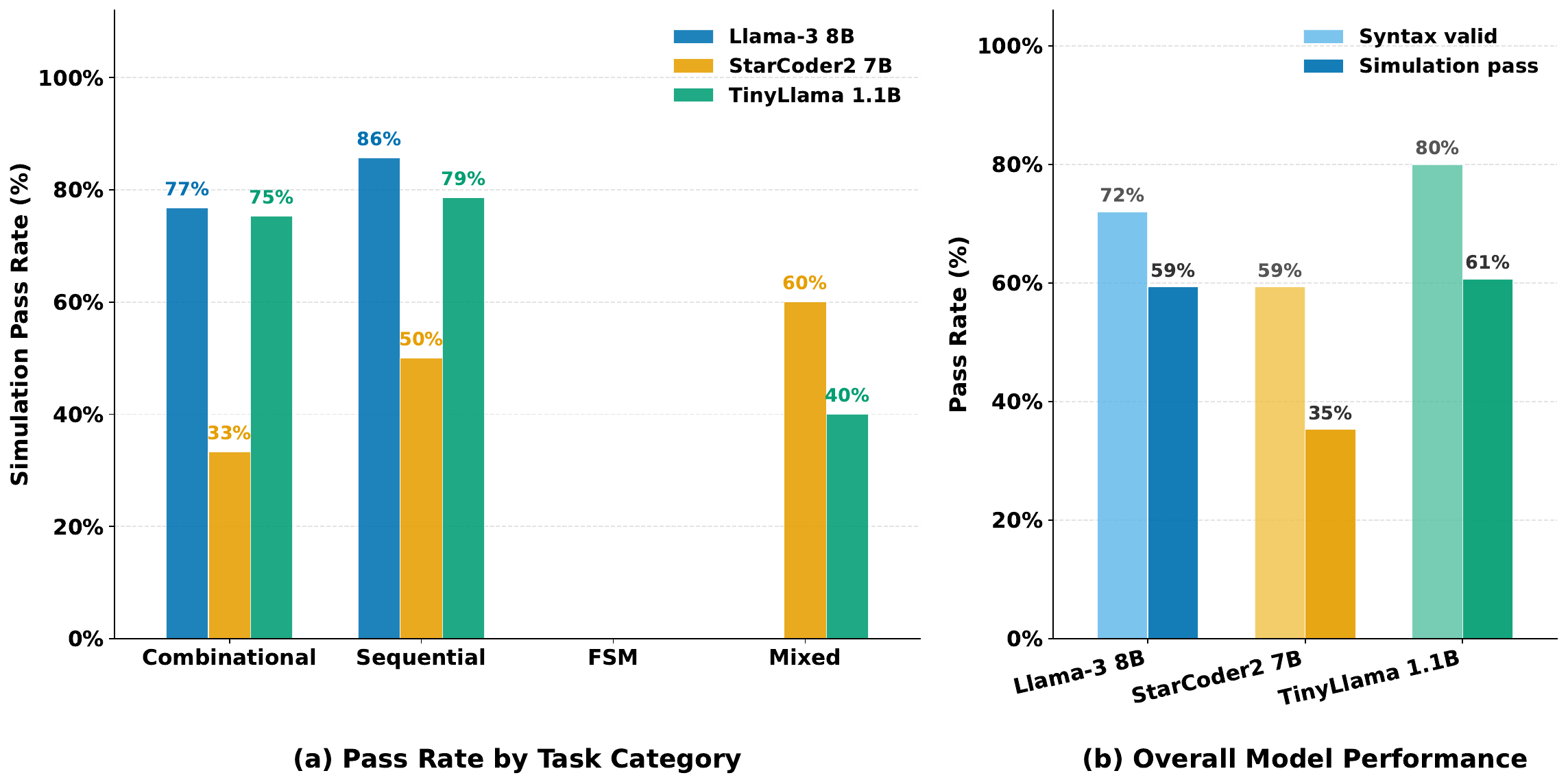}
    \caption{Benchmark 12 results: simulation pass rates by task category and overall model
    performance (syntax valid vs.\ simulation pass).}
    \label{fig:benchmark12}
\end{figure*}
\section{Evaluation Setup}

Three models span a wide capability range. Llama-3-8B-Instruct is a general-purpose instruction-tuned
model, accessed via Ollama. TinyLlama-1.1B, also via Ollama, is 7$\times$ smaller. StarCoder2-7B,
accessed via HuggingFace Transformers, is trained on large code corpora, but Verilog represents a
small fraction of that data. All models ran at temperature 0.0 with a 512-token limit. Each task is then repeated three times per model, with the results throughout being the mean $\pm\sigma$ across repetitions.

All the experiments were run on macOS (Apple Silicon) using Icarus Verilog 12.0, Verilator 5.038, and Yosys 0.58, which was installed via Homebrew. The cross-platform reproducibility is validated with a docker image pre-installing the full toolchain, and a mini benchmark (5 tasks $\times$ 3 models) confirms the execution is correct inside the container. To fully specify every run, all the runs (model, temperature, repetition count, and timeouts) are declared in a single \texttt{instruction.json} file.

\section{Results and Ablation Study}

The ablation study in Table~\ref{tab:benchmark_metrics} has demonstrated how each pipeline component contributes through improvements in syntax validity and simulation pass rate. Vanilla prompting produced a syntax validity of 0\%, constrained prompting (Benchmark~2) raised this syntax validity to 20\%, post-processing (Benchmark~3) further drives it to 60\%, the largest single-step
gain, larger than any difference between the models, timing instrumentation (Benchmark 4) held syntax validity at 63\%, and adding the third model (Benchmark 5) dipped it to 51.1\%; sequential normalization in Benchmark 6 reached 80\% syntax with a simulation of 73.3\%, but this was on a 5-task pilot where FSMs had not yet appeared.

Benchmark 7 then introduces 20 tasks, which include FSM and mixed types, which caused the syntax to fall to 29.4\%, showcasing clearly how much more difficult the FSM is than any other result. FSM template scaffolding showed some recovery in the 8th benchmark with an increment to 63.3\%. Phase 4 iterative
refinement then helps stabilize the performance at approximately 70\% syntax with a 49\% simulation on the full 50-task dataset (Benchmarks 9-10); Phase 5 improved the  structural correctness for the hardest task categories while holding the overall baseline. Benchmark 11 was mainly for the reproducibility check, running only Llama-3-8B on the benchmark 10 setup, causing it to produce a replicated 71.3\% syntax and 61.3\% simulation result while cutting the average generation time by 46\% (8.84~s to 4.77~s).

\begin{table}[!ht]
  \centering
  \caption{Per-Model Results for Benchmark 12 (mean\,$\pm$\,$\sigma$, 3 repetitions)}
  \label{tab:per_model_b12}
  \scriptsize
  \setlength{\tabcolsep}{3pt}
  \resizebox{\columnwidth}{!}{
    \begin{tabular}{|l|c|c|c|c|c|}
      \hline
      \textbf{Model} & \textbf{Syntax (\%)} & \textbf{Sim Pass (\%)} &
      \textbf{Synth (\%)} & \textbf{Avg Cells} & \textbf{Gen Time}\\
      \hline
      TinyLlama-1.1B & $80.0 \pm 3.5$ & $60.7 \pm 5.0$ & 75.3 & 5.6 & 3.72\,s\\
      Llama-3-8B     & $72.0 \pm 2.0$ & $59.3 \pm 1.2$ & 70.0 & 3.5 & 4.71\,s\\
      StarCoder2-7B  & $59.3 \pm 1.2$ & $35.3 \pm 3.1$ & 54.7 & 12.1 & 1.95\,s\\
      \hline
    \end{tabular}%
  }
\end{table}

Table~\ref{tab:per_model_b12} shows the $\sigma$ and Yosys synthesis results, where TinyLlama has led on the syntax validity with 80.0\%$\pm$3.5\% and has nearly matched Llama-3-8B when it came to functional correctness (60.7\%$\pm$5.0\%
vs.\ 59.3\%$\pm$1.2\%). Importantly, in this constrained and repair-assisted workflow, TinyLlama has achieved competitive results, suggesting that the workflow design may matter as much as the model scale for small RTL tasks~\cite{b8}, whereas StarCoder2 kept underperforming for both models despite the code specialization; also, its synthesizable outputs had the highest average cell count (12.1 vs.\ 5.6 and 3.5), which points to a structurally verbose generation rather than compact RTL.

Yosys synthesis had confirmed that 300 of the 450 Benchmark 12 runs (66.7\%) are synthesizable, and the sequential designs had the highest synthesis rate (85.7\%, avg.\ 5.0~cells), which was followed by combinational (67.1\%, 3.1~cells). FSM (43.1\%) and mixed (48.9\%) designs were lagging, with even the syntax-valid FSM outputs failing synthesis at a meaningful rate since the state encoding errors and missing \texttt{default} cases kept creating synthesis-blocking holes that the simulation stimulus alone does not reach. Mixed designs reflected unpredictable control-datapath partitioning by showing the highest cell count variance (6-88 cells), and Fig.~\ref{fig:benchmark12} breaks down the simulation pass rates by category and model.

\section{Failure Analysis: What Still Breaks?}
\begin{figure*}[!ht]
    \centering
    \includegraphics[width=\linewidth]{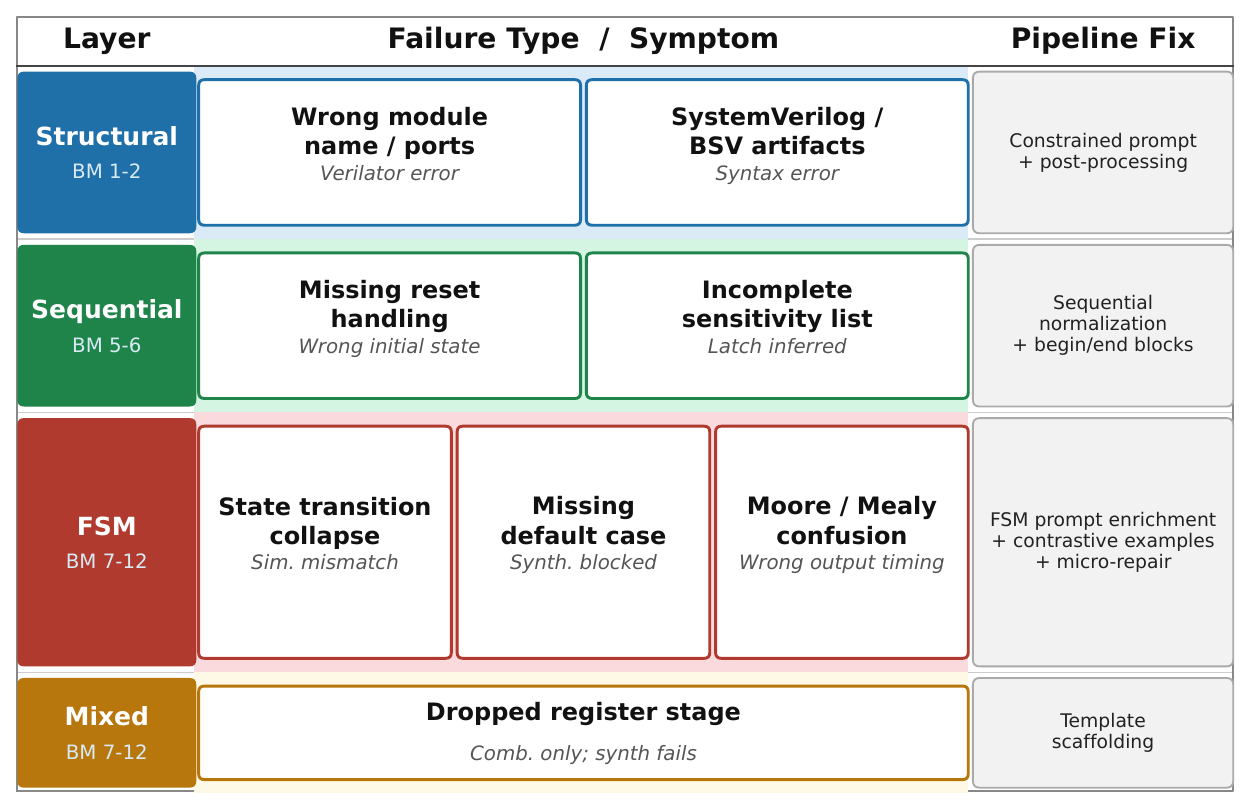}
    \caption{Failure taxonomy across the 12-benchmark progression. Each row shows a failure
    layer, the individual failure types with their observable symptom, and the pipeline
    component that addresses it. Structural failures respond fully to prompting and
    post-processing; FSM and mixed failures persist despite Phase~5 enhancements.}
    \label{fig:taxonomy}
\end{figure*}

The Fig.~\ref{fig:taxonomy} shows how the failures across the 12 benchmarks are clustered into four layers. Structural failures (Benchmarks 1-2) occur due to interface failures like wrong module names, missing port declarations, and SystemVerilog keywords where Verilog-2001 is required, and these have shown a good response to constrained prompting and post-processing, as they were essentially eliminated by Benchmark 4. Missing resets and latch inference from incomplete sensitivity lists were the primary causes of the sequential logic failures (Benchmarks 3–8), which are addressed by the template scaffolding. FSM and mixed failures persist throughout the Benchmark 12 despite Phase 5 enhancements because three distinct FSM failure modes recur across all the three models~\cite{b11}.

\textbf{Failure 1: State transition collapse.}
The most common FSM failure could be illustrated using the turnstile controller task, where the model copies a transition rather than inverting it, leading to a state that is reachable but never exits.

\begin{lstlisting}[style=verilogstyle, caption={Turnstile controller example illustrating state collapse. The generated FSM remains in the UNLOCKED state after a coin insertion and fails to return to the LOCKED state when a push event occurs.}]
// Generated: UNLOCKED never exits on push
always @(posedge clk or posedge rst)
  if (rst) state <= LOCKED;
  else case (state)
    LOCKED:   if (coin) state <= UNLOCKED;
    UNLOCKED: if (push) state <= UNLOCKED; // wrong
  endcase  // no default

// Reference
always @(posedge clk or posedge rst)
  if (rst) state <= LOCKED;
  else case (state)
    LOCKED:   if (coin) state <= UNLOCKED;
    UNLOCKED: if (push) state <= LOCKED;
    default:  state <= LOCKED;
  endcase
\end{lstlisting}

At cycle 4, \texttt{WaveformAnalyzer} detects the divergence, while Yosys formal equivalence confirms a mismatch in the state space. However, even after receiving this feedback, the model continues to regenerate FSMs with similar structural errors in subsequent iterations, indicating that the feedback is not sufficiently precise to guide the repair process.

\textbf{Failure 2: Moore/Mealy confusion.}
A timing-related error was observed in the sequence detector tasks, where the model implemented the output logic as combinational despite the specification requiring a registered (Moore-style) output. As a result, the generated output is asserted one clock cycle earlier than expected. In edge-case scenarios, the design fails even though it might pass simple test vectors. Waveform comparison, when the generated signal regularly lags the reference output by one clock cycle, reveals this problem, which is not picked up by linting or basic simulation stimuli.

\textbf{Failure 3: Missing \texttt{default} blocking synthesis.} The model at the Verilog level continued to generate a combinational always block for next-state logic with full case coverage in traffic light FSM jobs, but it omitted the \texttt{default} branch. This was a consistent synthesis failure. The inclusion of synthesis checks to the pipeline, with the micro-repair pass introduced in Phase 5 expressly targeting this pattern, was inspired by the fact that this pattern passes the simulation (the testbench never exercises an unencoded state) but fails the synthesis. Mixed designs fail by a fourth mechanism, where models drop the registered output stage entirely and implement only the combinational path of a control-datapath design, leading the result to pass combinational test vectors but breaking the logic in downstream sequential. Crucially, all the four patterns share a common root cause in which the models lack sufficient structural reasoning to maintain a consistent hardware abstraction (state graph, registered output, complete case coverage) across the generated text of 30-50 lines.

\section{Lessons for EDA Practitioners}

This study highlights five key points for organizations evaluating LLM-assisted RTL operations:

\textbf{1. Prompt precision matters more than model size.} Injecting the precise module names, port lists, and Verilog-2001 rules produced the biggest single improvement across all 12 benchmarks; however, the difference between constrained and unconstrained prompting was greater than the difference between any two models tested, so start there before scaling up the model.

\textbf{2. Post-processing is not optional.} It was observed that all the three models hallucinate BSV artifacts, use SystemVerilog keywords, and generate code outside module boundaries with measurable frequency even with the inclusion of constrained prompts. Regex-based cleanup handles any kind of common cases, and
the structural issues missed by regex are caught by AST-level repair. Therefore, 30–40\% of otherwise accurate outputs are uncompilable if either of the layers is skipped.

\textbf{3. Iterative feedback works for syntax, but not for FSM logic.} The syntax corrections are reliably driven by the compiler error feedback, with the simulation mismatch feedback being effective for combinational
tasks, but even five additional repair iterations do not significantly improve the functional correctness on the most difficult jobs because the feedback signal for FSM state transition errors is too coarse.

\textbf{4. Match task complexity to the right model.} TinyLlama-1.1B matched Llama-3-8B when it came to combinational and standard sequential
tasks, and for FSMs and mixed designs, the larger model tended to perform better, showcasing how small models are appropriate for leaf modules and simple datapath elements but stateful control logic warrants a larger model or human review.

\textbf{5. Reproducibility demands infrastructure.} Run-to-run variance
is measured even at temperature 0.0 (Table~\ref{tab:per_model_b12}). Three-repetition averaging, containerized toolchains, and structured result recording imply these are the minimum standards, not academic overhead, that are required before trusting LLM-generated RTL in a practical design cycle.

\section{Limitations and Future Directions}
This study has three defined boundaries, beginning with the dataset that covers single-module, fully specified, and deterministically testable designs, but hierarchical RTL, bus protocols, and multi-clock-domain logic are absent because this platform does not simulate task decomposition, incremental verification, or human-in-the-loop review at decision points, all of which are necessary for production RTL generation at that high scale, but expanding toward those patterns could be the primary direction for future work, along with larger task sets that include IP-level complexity.

Additionally, post-synthesis quality metrics (area, timing, power) are not measured because for hardware use, synthesizability is a prerequisite rather than a sufficient one, and whether LLM-generated RTL can meet real timing constraints under process corners still remains an open question; thus, the planned next step would be extending the Yosys evaluation to technology-mapped netlists and static timing~\cite{b3,b12}.

When it comes to threats to validity, the reference testbench in this study defines the  functional correctness. There may be overreported pass rates caused by tasks where the testbench applies limited stimulus coverage, along with run-to-run variance at temperature 0.0 (documented in the $\sigma$ column of Table~\ref{tab:per_model_b12}) arising from hardware-level nondeterminism in Apple Silicon inference; therefore, all results are three-repetition means. Although not tested for additional models, operating systems, or toolchain versions, the Docker image reduces reliance on the toolchain.

\section{Conclusion}

Guarded generations like constrained prompting, structured post-processing, semantic-aware iterative repair, and real-toolchain validation have changed LLM-generated Verilog that was wholly non-compilable into functionally accurate on the majority of combinational and sequential RTL jobs. Yet FSMs remain a
hard case, where even with Phase 5 enhancements, there are reasoning gaps being exposed by complex stateful designs that weren't reliably closed by any current 1-8B open-source model. One noteworthy discovery is that, in this workflow, TinyLlama, a 1.1B model, led on syntax validity and nearly matched an 8B model on simulation. This finding is consistent with the idea that model scale and workflow infrastructure are important for constrained RTL tasks. The platform and dataset are open-source, and every result in this paper is reproducible from a single \texttt{instruction.json}.



\balance
\vspace{11pt}

\begin{IEEEbiography}[{\includegraphics[width=1in,height=1.25in,clip,keepaspectratio]{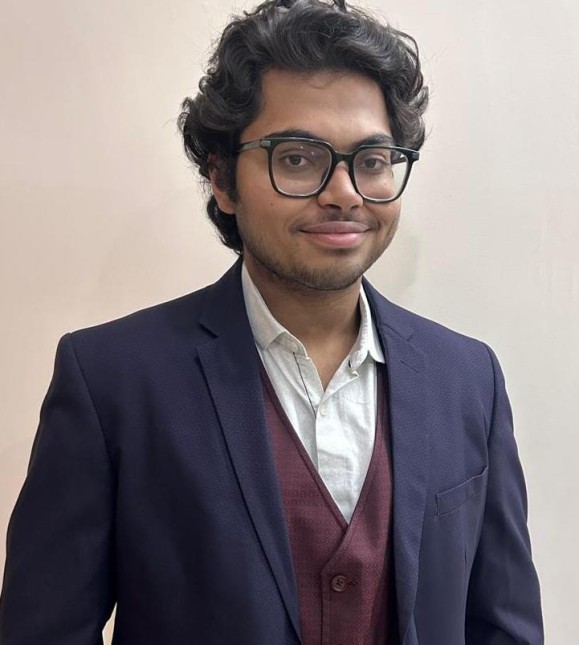}}]{Angshuman Chakravertty} is currently pursuing the B.Tech.\ degree in computer science and engineering (data science) at the School of Technology Management and Engineering, NMIMS, Hyderabad, India (expected 2027). His research interests include artificial intelligence and machine learning, with a focus on their applications to real-world engineering problems.
\end{IEEEbiography}
\vskip -1\baselineskip plus -1fil
\begin{IEEEbiography}
[{\includegraphics[width=1in,height=1.25in,clip,keepaspectratio]{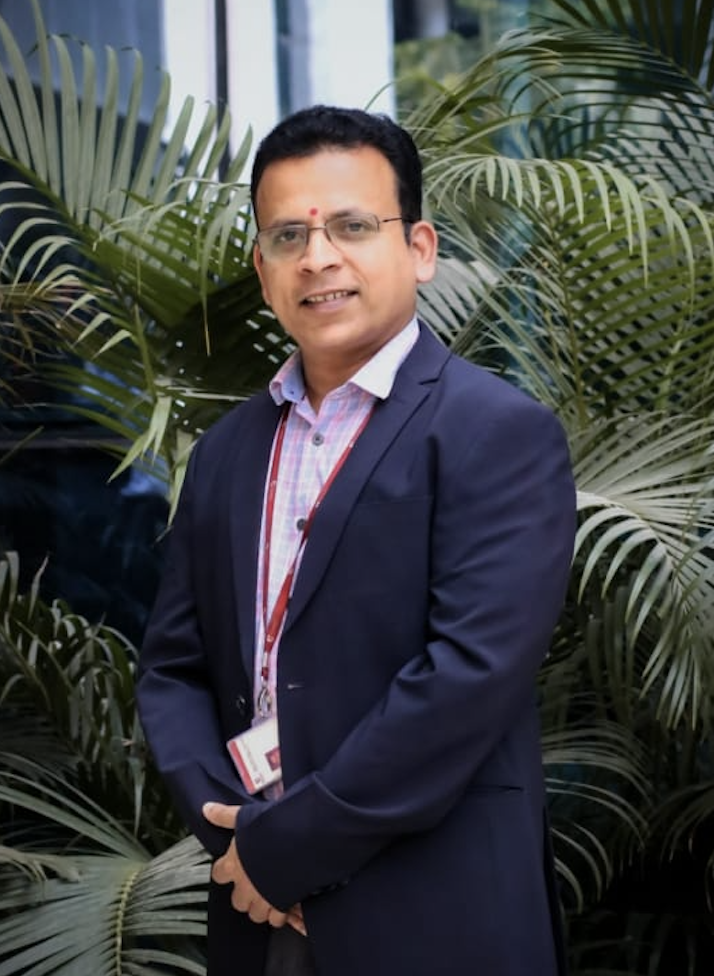}}]
{Rahul Koshti}
received the M.E. degree with a specialization in digital communication and the Ph.D. degree in wireless communication. He has around 20 years of experience in education and administration and is a certified professional in educational leadership. He is currently with the School of Technology Management and Engineering, NMIMS, Hyderabad, India, where he also heads the Department of Placements and Corporate Relations. Over his career he has held roles spanning training and placements, corporate relations, examinations, and accreditation, gaining extensive experience in institutional administration, policy framing, and the establishment of new academic ventures.
His research interests include wireless communication, antenna design, and cognitive-radio-based wireless networks. He has published research papers, patents, and articles in reputed IEEE conferences and journals indexed in Scopus, Springer Nature, and IEEE Xplore, along with book chapters with CRC Press, Taylor \& Francis. He is an active member of the IEEE and the ISOC Hyderabad chapter and serves as counsellor of the IEEE Student Branch. He has delivered invited expert talks on wireless communication, IoT, and NAAC accreditation, and has organized several national and international conferences.
\end{IEEEbiography}
\vskip -1\baselineskip plus -1fil
\begin{IEEEbiography}[{\includegraphics[width=1in,height=1.25in,clip,keepaspectratio]{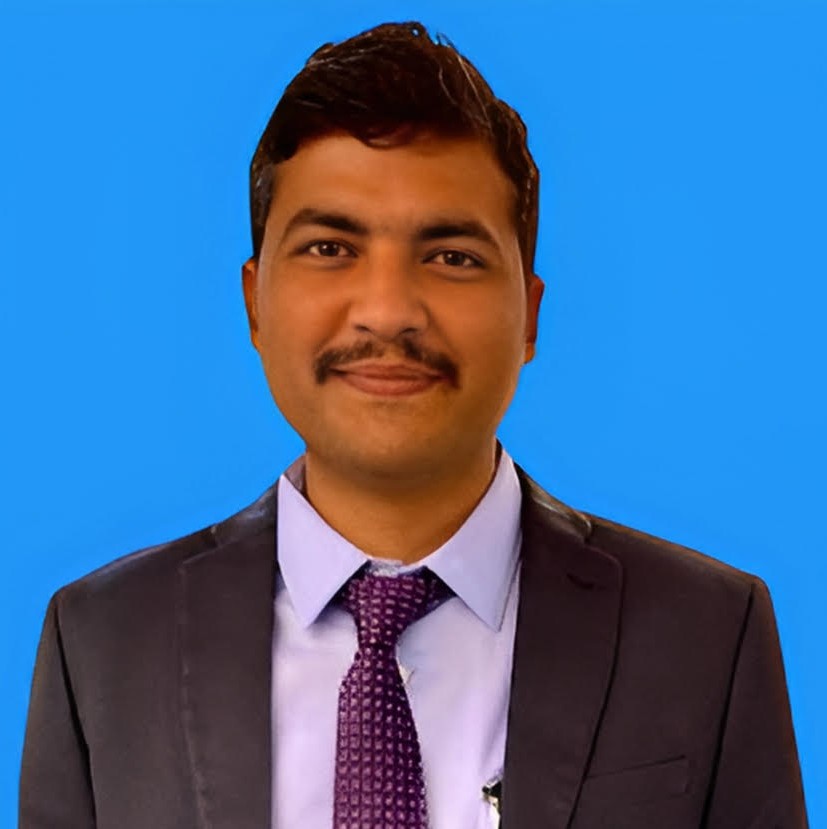}}]{Buddhi Prakash Sharma}
received a B.Tech degree from ICFAI University, Dehradun, India, in 2009, an M.E. degree from Punjab University, Chandigarh, India, in 2017, and completed a Ph.D. degree from Birla Institute of Technology \& Science, Pilani, India in 2026. From 2009 to 2021, he was an Assistant Professor in academic institutes. His research interests include Analog and mixed signal circuit design, VLSI Design, and Image processing. He is currently focusing his research work on Data Converters for IoT applications.
\end{IEEEbiography}

\vskip -1\baselineskip plus -1fil
\begin{IEEEbiography}[{\includegraphics[width=1in,height=1.25in, clip,keepaspectratio]{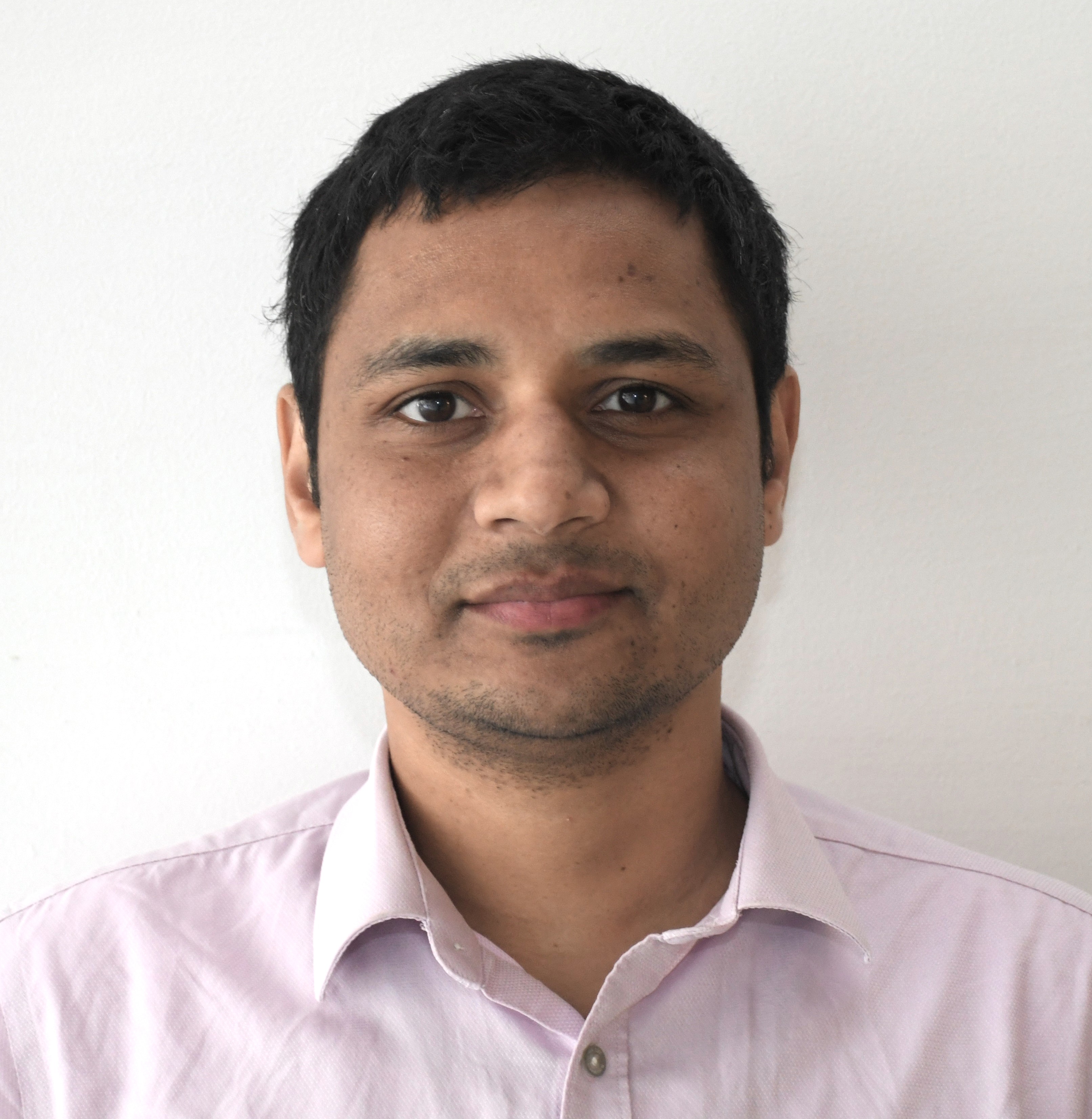}}]{Vinay Chamola} received the B.E. degree in electrical and electronics engineering and master’s degree in communication engineering from the Birla Institute of Technology and Science, Pilani, India, in 2010 and 2013, respectively. He received his Ph.D. in electrical and computer engineering from the National University of Singapore, Singapore, in 2016. In 2015, he was a Visiting Researcher with the Autonomous Networks Research Group (ANRG), University of Southern California, Los Angeles, CA, USA. He also worked as a post-doctoral research fellow at the National University of Singapore, Singapore. 
He is an Associate Professor with the Department of Electrical and Electronics Engineering, BITS-Pilani, Pilani, where he heads the Internet of Things Research Group / Lab. His research interests include IoT Security, Blockchain, UAVs, VANETs, 5G, and Healthcare. He serves as an Area Editor for the Ad Hoc Networks Journal, Elsevier, and the IEEE Internet of Things Magazine. He is also an Associate Editor in the IEEE Transactions on Intelligent Transportation Systems, IEEE Networking Letters, IEEE Consumer Electronics magazine, IET Quantum Communications, IET Networks, and several other journals. He serves as co-chair of various reputed workshops like IEEE Globecom Workshop 2021, IEEE INFOCOM 2022 workshop, IEEE ANTS 2021, and IEEE ICIAfS 2021, to name a few. He is listed in the World’s Top 2\% Scientists identified by Stanford University. He is the co-founder and President of a healthcare startup, Medsupervision Pvt. Ltd. He is a senior member of the IEEE and a Fellow of the IET.

\end{IEEEbiography}


\begin{thebibliography}{12}

\bibitem{b2}
S.~Thakur, B.~Ahmad, Z.~Fan, H.~Pearce, B.~Tan, R.~Karri, B.~Dolan-Gavitt, and S.~Garg,
``Benchmarking large language models for automated Verilog RTL code generation,'' in
\emph{Proc.\ 2023 Design, Autom.\ \& Test in Europe Conf.\ (DATE)}, Antwerp, Belgium,
Apr.~2023, pp.~1--6.

\bibitem{b1}
S.~Thakur \emph{et al.}, ``VeriGen: A large language model for Verilog code generation,''
\emph{ACM Trans.\ Design Autom.\ Electron.\ Syst.}, vol.~29, no.~3, pp.~1--31, Apr.~2024.

\bibitem{b6}
M.~Liu, N.~Pinckney, B.~Khailany, and H.~Ren, ``VerilogEval: Evaluating large language
models for Verilog code generation,'' in \emph{2023 IEEE/ACM Int.\ Conf.\ on Computer-Aided
Design (ICCAD)}, IEEE, Nov.~2023, pp.~1--8.

\bibitem{b4}
M.~Gao \emph{et al.}, ``AutoVCoder: A systematic framework for automated Verilog code generation
using LLMs,'' in \emph{2024 IEEE 42nd Int.\ Conf.\ on Computer Design (ICCD)}, IEEE, Nov.~2024,
pp.~162--169.

\bibitem{b5}
R.~Qiu \emph{et al.}, ``AutoBench: Automatic testbench generation and evaluation using LLMs
for HDL design,'' in \emph{Proc.\ 2024 ACM/IEEE Int.\ Symp.\ on Machine Learning for CAD
(MLCAD~'24)}, Salt Lake City, UT, USA: ACM, Sep.~2024, pp.~1--10.

\bibitem{b7}
Y.~Yang \emph{et al.}, ``HaVen: Hallucination-mitigated LLM for Verilog code generation
aligned with HDL engineers,'' in \emph{2025 Design, Autom.\ \& Test in Europe Conf.\ (DATE)},
IEEE, 2025, pp.~1--7.

\bibitem{b9}
Z.~Bi \emph{et al.}, ``Iterative refinement of project-level code context for precise code
generation with compiler feedback,'' in \emph{Findings of the Assoc.\ for Computational
Linguistics}, Jan.~2024, pp.~2336--2353.

\bibitem{b10}
N.~Wang \emph{et al.}, ``VeriDebug: A unified LLM for Verilog debugging via contrastive
embedding and guided correction,'' in \emph{Proc.\ 2025 IEEE Int.\ Conf.\ on LLM-Aided
Design (ICLAD)}, IEEE, 2025, pp.~61--67.

\bibitem{b8}
P.~Yubeaton \emph{et al.}, ``VeriThoughts: Enabling automated Verilog code generation using
reasoning and formal verification,'' May~2025, arXiv preprint.

\bibitem{b11}
H.~Qin \emph{et al.}, ``ReasoningV: Efficient Verilog code generation with adaptive hybrid
reasoning model,'' Apr.~2025, arXiv preprint.

\bibitem{b3}
M.~Abdelatty, J.~Ma, and S.~Reda, ``MetRex: A benchmark for Verilog code metric reasoning
using LLMs,'' in \emph{Proc.\ 30th Asia and South Pacific Design Autom.\ Conf.\ (ASPDAC~'25)},
New York, NY, USA: ACM, 2025, pp.~995--1001.

\bibitem{b12}
K.~Thorat \emph{et al.}, ``LLM-VeriPPA: Power, performance, and area optimization aware
Verilog code generation with large language models,'' in \emph{2025 IEEE/ACM Int.\ Conf.\ on
Machine Learning for Electronic Design Autom.\ (MLCAD)}, Sep.~2025, pp.~1--7.

\end{thebibliography}
\end{document}